\documentclass[aip,prl,twocolumn]{revtex4}
\usepackage{graphics,graphicx,color}
\begin{document}

\title[]{Spin wave damping arising from phase coexistence below $T_c$ in colossal magnetoresistive La$_{0.7}$Ca$_{0.3}$MnO$_3$}
\author{Joel~S.~Helton$^{1,2,\ast}$, Susumu~K.~Jones$^{1}$, Daniel~Parshall$^{2}$, Matthew~B.~Stone$^{3}$, Dmitry~A.~Shulyatev$^{4}$, and Jeffrey~W.~Lynn$^{2,\dag}$}

\address{$^{1}$Department of Physics, United States Naval Academy, Annapolis, MD 21402, USA}
\address{$^{2}$NIST Center for Neutron Research, National Institute of Standards and Technology, Gaithersburg, MD 20899, USA}
\address{$^{3}$Quantum Condensed Matter Division, Oak Ridge National Laboratory, Oak Ridge, TN 37831, USA}
\address{$^{4}$National University of Science and Technology ``MISiS", Moscow 119991, Russia}

\begin{abstract}
While the spin dynamics of La$_{0.7}$Ca$_{0.3}$MnO$_3$ in the ferromagnetic phase are known to be unconventional, previous measurements have yielded contradictory results regarding the damping of spin wave excitations.  Neutron spectroscopy measurements on a sample with a transition temperature of $T_c$=257~K, higher than most single crystals, unambiguously reveal an anomalous increase in spin wave damping for excitations approaching the Brillouin zone boundary along the [$100$] direction that cannot be explained as an artifact due to a noninteracting phonon branch.  Spin waves throughout the ($HK0$)~plane display a common trend where the spin wave damping is dependent upon the excitation energy, increasing for energies above roughly 15~meV and reaching a full width at half maximum of at least 20~meV.  The results are consistent with a model of intrinsic spatial inhomogeneity with phase separated regions approximately 18~{\AA} in size persisting over a large range of temperatures below $T_c$.

\end{abstract}
\maketitle

\section{I. Introduction}

Hole-doped perovskite manganites of the form $L_{1-x}A_x$MnO$_3$ (where $L$ is a trivalent lanthanide ion and $A$ is a divalent alkaline earth ion)\cite{Wollan1955,Raveau1998} display a range of unusual physical properties arising from intrinsic spatial inhomogeneities such as nanoscale phase separation\cite{Dagotto2003,Moreo1999} and polarons\cite{deTeresa,Adams2000}.  While much of the interest in these materials has been concentrated on the colossal magnetoresistance (CMR) observed at the combined ferromagnetic and metal-insulator phase transition, the spin dynamics observed in these compounds in the ferromagnetic regime are also unconventional and cannot be fully explained by a simple model of a nearest-neighbor Heisenberg ferromagnet\cite{Lynn2000,Zhang2007}. The spin wave dispersion is significantly softened near the zone boundary when compared to the low-$q$ spin wave stiffness\cite{Hwang1998,FernandezBaca1998,Dai2000} and is often fit to a phenomenological model of first- and fourth-nearest-neighbor ferromagnetic interactions $J_1$ and $J_4$\cite{Ye2007}.  Significant broadening of the spin wave excitations has been observed in several compounds\cite{Hwang1998,VasiliuDoloc1998prb}, particularly for large-$q$ excitations approaching the Brillouin zone boundary.  These unconventional spin behaviors are generally more pronounced in samples near optimal doping ($x\approx0.3$) and in compounds with a lower $T_c$ such as La$_{1-x}$Ca$_{x}$MnO$_3$ (LCMO) or Nd$_{1-x}$Sr$_{x}$MnO$_3$\cite{Ye2006}.  These trends point to a connection between unconventional spin behaviors and colossal magnetoresistance, as a more pronounced CMR effect is generally present in these same materials.

\begin{figure}
\centering
\includegraphics[width=8.3cm]{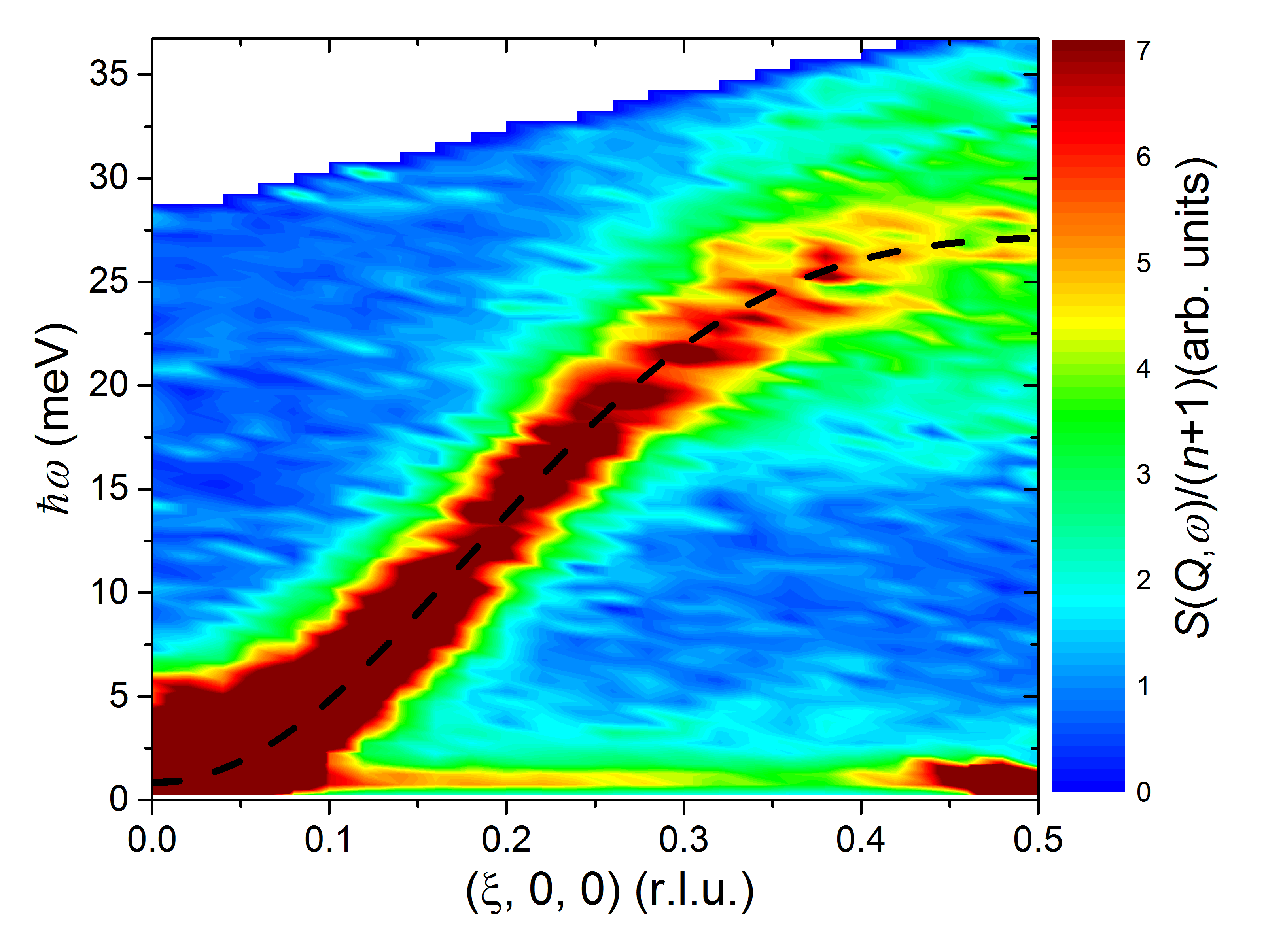} \vspace{-5mm}
\caption{Observed $S(\vec{q},\omega)/(n+1)$ for spin wave excitations along the [100] direction, measured at $T$=100~K with $E_i$=50~meV.  The dashed line is the dispersion curve.} \vspace{-2mm}
\label{Fig1}
\end{figure}

\begin{figure*}
\centering
\includegraphics[width=16.0cm]{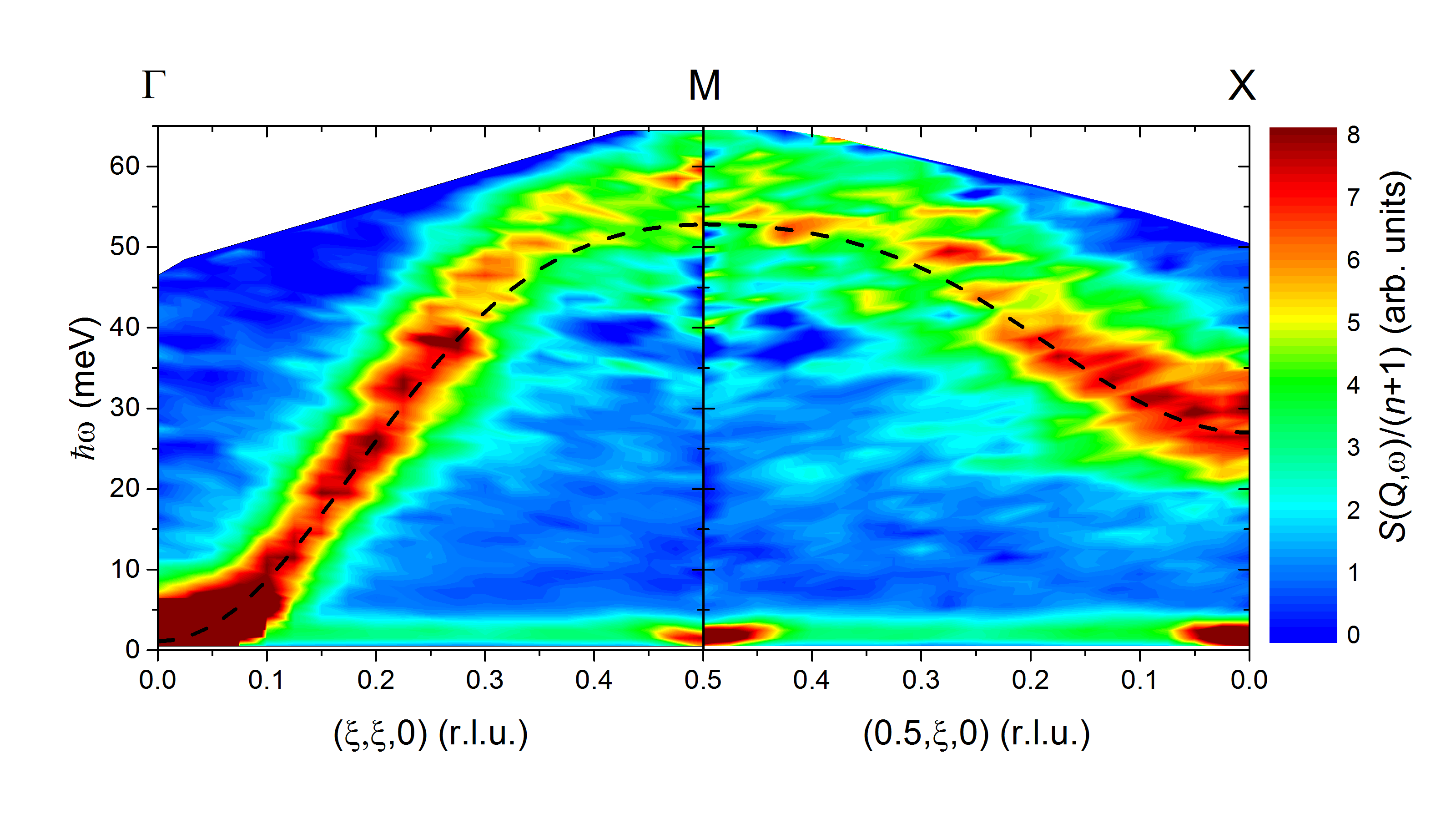} \vspace{-10mm}
\caption{Lower resolution $S(\vec{q},\omega)/(n+1)$ data for spin wave excitations at $\vec{q}=(\xi,\xi,$0) and $\vec{q}=(0.5,\xi,$0), measured at $T$=100~K with $E_i$=120~meV.} \vspace{-2mm}
\label{Fig2}
\end{figure*}

La$_{0.7}$Ca$_{0.3}$MnO$_3$ is a particularly important compound to understand.  The observed CMR effect is significant\cite{Jin1994}.  La$^{3+}$ is nonmagnetic, and the ionic radii of La$^{3+}$ and Ca$^{2+}$ are quite similar so that $\sigma_A^2$, the variance in $A$-site ionic radius and a common measure of structural disorder, is relatively small and all evidence indicates that the $A$ site is randomly occupied.  The ferromagnetic transition is also unconventional.  At temperatures approaching $T_c$, spectral weight is shifted from spin wave excitations into a quasielastic spin-diffusive component, and the spin wave stiffness renormalizes to a softer value but does not vanish at the weakly first-order phase transition\cite{Lynn1996,Adams2004}.  Short-range ferromagnetic spin correlations, taking the form of ridges of strong scattering running along [100]-type directions in momentum space, are observed near and above $T_c$\cite{Helton2012}; a simple model associating this scattering component with the magnetic part of diffuse polarons qualitatively describes the data\cite{Helton2014}.

The current literature contains several conflicting results regarding spin wave broadening in $x$=0.3~LCMO.  Dai, \emph{et al.}\cite{Dai2000} initially reported that at $T$=10~K the spin wave damping at $\vec{Q}=$($0,0,1+\xi$) increased anomalously for $\xi \geq 0.3$.  This position in momentum space also corresponds to a crossing between the spin wave branch and an optical phonon, so the broadening was attributed to strong magnetoelastic coupling. However, experiments on manganite materials have not found evidence of significant coupling at the crossing between the spin wave and phonon branches\cite{Moudden1998,Moussa2007,FernandezBaca2006}. The lattice and magnetic degrees of freedom are however coupled in the polaron regime above $T_c$\cite{Adams2001}, and in a variety of manganite compounds the local distortions in the crystallographic position of oxygen atoms have been found to increase at $T_c$\cite{Radaelli,Weber} while acoustic phonons in La$_{0.7}$Sr$_{0.3}$MnO$_3$ soften and broaden when heated through $T_c$\cite{Maschek}.  A later work by Ye \emph{et al.}\cite{Ye2007} reported significantly damped excitations, with $\Gamma$ (the full width at half maximum, FWHM) reaching 20~meV along the [$110$] direction and 60~meV along the [$111$] direction, but no abrupt change in broadening.  An additional work by Moussa \emph{et al.}\cite{Moussa2007} reported a similar broadening in raw data but analyzed the data as consisting of noninteracting spin wave and phonon branches that lie next to each other.  In this analysis the spin waves were claimed to neither mix with the phonon branch nor damp anomalously, with $\Gamma\approx0.24\hbar\omega$ for excitations along the $[001]$ direction.  Polarized neutron scattering was used to demonstrate that the broad linewidths observed in spin wave excitations at $\vec{Q}=(\xi,\xi,2)$ with $\xi=0.3$ to 0.5 are intrinsic to the magnetic scattering cross section\cite{FernandezBaca2006}.  In this paper we will show an abrupt increase in spin wave broadening for excitations at energies exceeding $\approx$15~meV.  While there is apparently sample dependence in measurements of spin wave damping the sample used in this study displays a transition temperature of $T_c$=257~K, higher than that reported for most other single crystals of LCMO, indicating close to optimal doping.

\section{II. Experiment and Results}

Neutron spectroscopy measurements were performed on a 1.5~g single crystal grown by the floating zone method\cite{Shulyatev2002}.  The crystal structure is an orthorhombic perovskite, but given the presence of multiple crystallographic domains the sample is indexed in the pseudocubic notation with $a$=3.9~{\AA}.  The inelastic scattering spectrum was measured at $T$=100~K using the ARCS time-of-flight spectrometer\cite{Abernathy2012} at the Spallation Neutron Source.  Data were taken at incident neutron energies of 50~meV and 120~meV; these settings provided energy resolutions (FWHM) at the elastic line of 1.7~meV and 3.8~meV, respectively. The sample was mounted in the $(HK0)$ plane and data were measured as the sample was rotated in 1$^\circ$ increments through an angular range of 76$^\circ$ (for the $E_i$=50~meV data) or 68$^\circ$ (for the $E_i$=120~meV data). ARCS data in this paper referenced by the reduced wave vector $\vec{q}$ were obtained by averaging momentum space positions within the ($HK0$)~plane surrounding the ($1,0,0$) and equivalent Bragg peaks.  Color plot data are averages over $\pm$0.16~{\AA}$^{-1}$ perpendicular to the plane ($-0.1\leq L \leq 0.1$) and $\pm$0.08~{\AA}$^{-1}$ in the transverse momentum space direction within the plane.  Reported widths are fits to data averaging over $\pm$0.10~{\AA}$^{-1}$ in both perpendicular momentum space directions ($E_i$=50~meV data) or $\pm$0.16~{\AA}$^{-1}$ perpendicular to the plane and $\pm$0.08~{\AA}$^{-1}$ in the transverse momentum space direction within the plane ($E_i$=120~meV data).  Further measurements were performed using the BT-7 triple-axis spectrometer\cite{Lynn2012} at the NIST Center for Neutron Research.  Higher-resolution measurements at $T$=100~K using  BT-7 were performed with $E_i$=13.7~meV and 25$^\prime$-10$^\prime$-10$^\prime$-25$^\prime$ collimators [for measurements at $\vec{Q}$=(0.95, 0, 0) and (1.1, 0 , 0)] or with $E_f$=13.7~meV and 25$^\prime$-10$^\prime$-25$^\prime$-25$^\prime$ collimators [for measurements at $\vec{Q}$=(1.14, 0, 0)].  Lower-resolution measurements on BT-7 were performed for $\vec{Q}$=(1.2, 0, 0) at $T$=100~K as well as $\vec{Q}$=(1.4, 0, 0) at a variety of temperatures with $E_f$=14.7~meV and Open-50$^\prime$-50$^\prime$-120$^\prime$ collimators. Error bars and uncertainties throughout this paper are statistical in nature and represent one standard deviation.

\begin{figure}
\centering
\includegraphics[width=8.3cm]{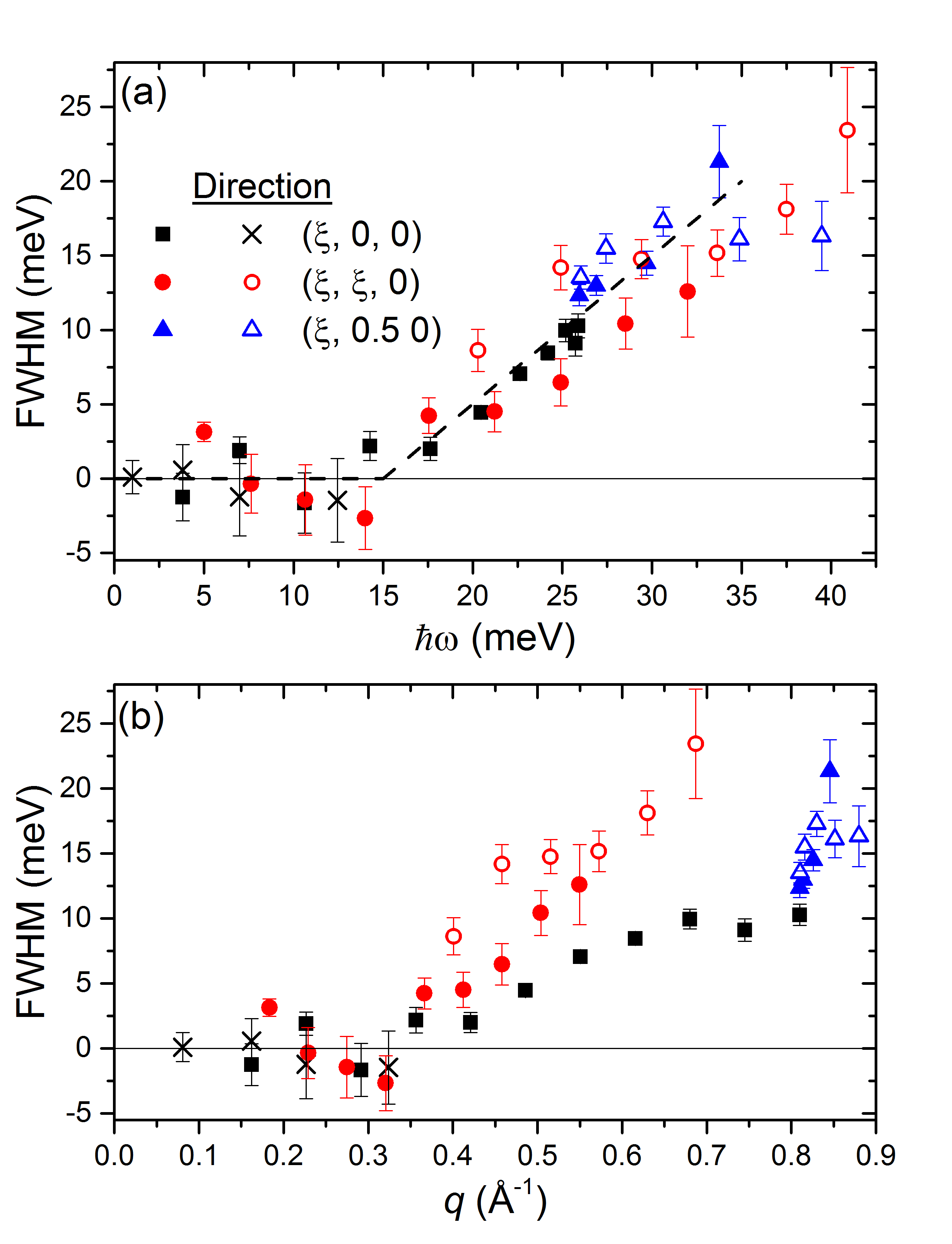} \vspace{-5mm}
\caption{(a) Intrinsic spin wave full-width at half-maximum (FWHM) as a function of excitation energy for three high-symmetry directions, all measured at $T$=100~K.  The dashed line is a guide to the eye.  (b) Intrinsic spin wave widths as a function of the reduced wave vector magnitude.  Widths shown with solid symbols were obtained by fitting the $E_i$=50~meV data from ARCS. Widths shown with open symbols were obtained by fitting the coarser resolution $E_i$=120~meV data.  Widths shown with an $\times$ were obtained by fitting data from BT-7.} \vspace{-2mm}
\label{Fig3}
\end{figure}

Figure~\ref{Fig1} shows spin wave excitations along the $\Gamma-X$ direction [$\vec{q}=(\xi,0,0$)], measured at $T$=100~K on the ARCS spectrometer with $E_i$=50~meV.  The color plot displays $S(\vec{Q},\omega)/(n+1)$ (where $n$ is the Bose occupation factor). While the linewidths are resolution limited for small reduced wave vectors, the spin wave excitations abruptly exhibit monotonically increasing linewidths for reduced wave vectors with $\xi\gtrsim0.25$.  While an optical phonon mode is known to be present at energies near 20~meV, it is not discernible in this data due to the small dynamic structure factor for phonon scattering at low $Q$. These results are similar to earlier reports on magnon broadening at low temperatures\cite{Dai2000}. 

The spin wave dispersion in lower $T_c$ manganites is frequently fit to a model with both first- and fourth-nearest-neighbor ferromagnetic interactions $J_1$ and $J_4$\cite{Hwang1998}, though this model tends to underestimate the spin wave energy for the largest excitation energies\cite{Ye2007}. The dispersion curve for this compound at $\vec{q}=(\xi,0,0$) is given by $\hbar\omega=2S[|J_1|(1-\textrm{cos}(2\pi\xi))+|J_4|(1-\textrm{cos}(4\pi\xi))]$ where fitted values for the ferromagnetic interactions are given by $S|J_1|$=6.47$\pm$0.15~meV and $J_4/J_1$=0.15$\pm$0.02. Dashed lines in color plots represent the peak values of the convolution of the dispersion curve with the instrumental resolution.

\begin{figure}
\centering
\includegraphics[width=9.3cm]{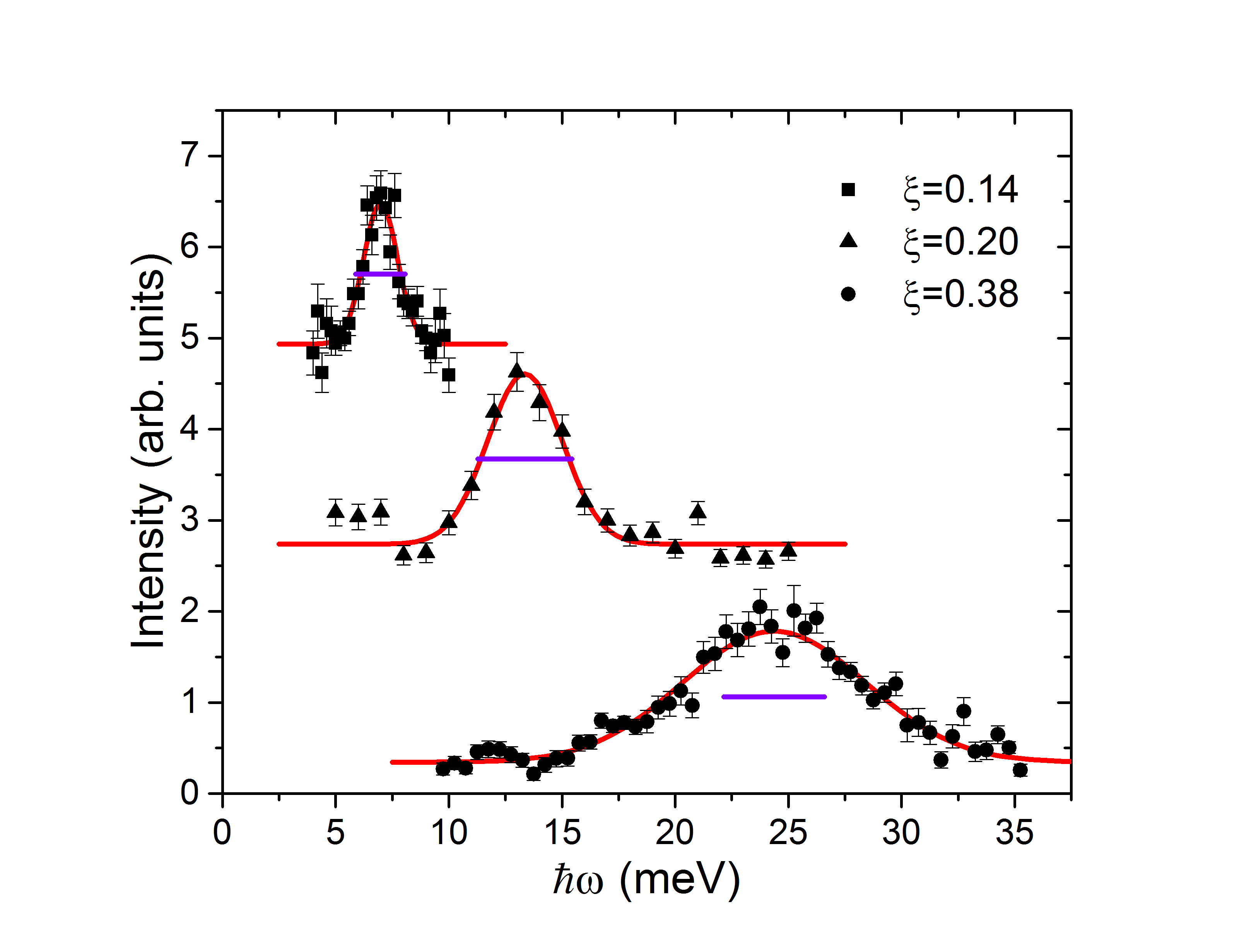} \vspace{-5mm}
\caption{Example spin wave excitations at $T$=100~K for $\vec{q}=$($\xi$, 0, 0) with $\xi$=0.14, 0.20, and 0.38. The $\xi$=0.20 and $\xi$=0.14 data were offset upward by, respectively, two and four arbitrary units of intensity.  The curves through the data are fits.  The horizontal lines represent the instrumental resolution (FWHM).} \vspace{-2mm}
\label{Fig4}
\end{figure}

\begin{figure*}
\centering
\includegraphics[width=18.5cm]{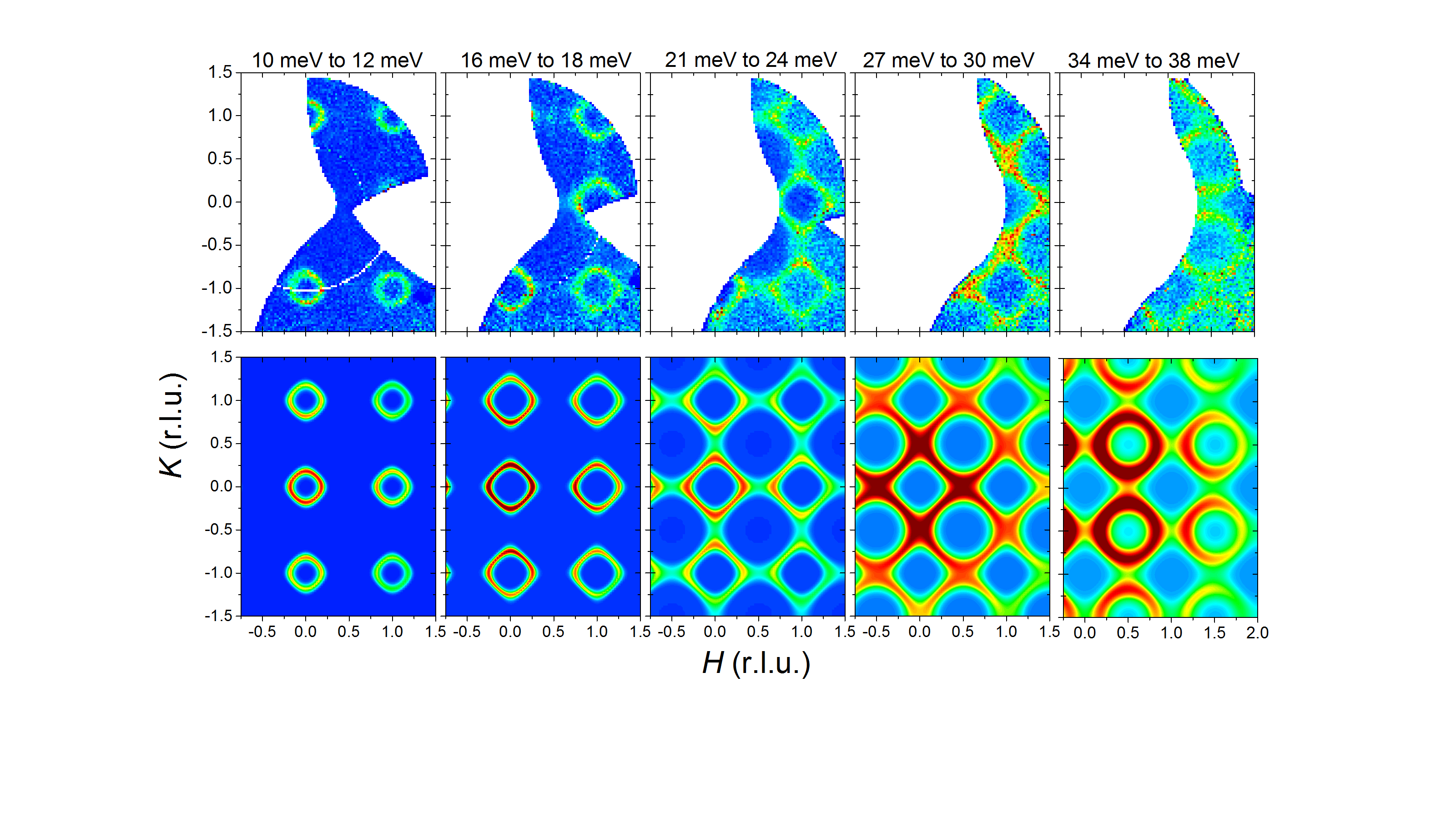} \vspace{-25mm}
\caption{Data (top row) and simulation (bottom row) for $S(\vec{Q},\omega)$, in arbitrary intensity units, over five energy binning ranges.  All plots have been binned over $\pm$0.16~{\AA}$^{-1}$ perpendicular to the plane ($-0.1\leq L \leq 0.1$). The color scales are consistent for data and simulation over the same energy range, but for clarity vary between energy ranges.} \vspace{-2mm}
\label{Fig5}
\end{figure*}

A recent work\cite{Moussa2007} measuring spin waves in $x$=0.3 LCMO reported raw data showing a comparable anomalous broadening at high $q$, but this broadening was analyzed as an artifact of noninteracting phonon modes lying near the spin wave branch. The spin wave damping described here cannot be similarly explained.  While phonon branches could be observed in cuts through the data in higher-$Q$ Brillouin zones, no phonon intensities are discernible in the lower-$Q$ data used for these fits.  Further, the intensity of phonon modes near the ($2,0,0$) Bragg position were determined and used to estimate an upper bound to the phonon intensity near the ($1,0,0$) position; including a phonon branch with this intensity did not alter the fit results.  These differences may indicate some sample dependence of anomalous magnon damping in $x$=0.3~LCMO.  The sample used here has a very high transition temperature ($T_c$=257~K) compared to most other single crystal LCMO samples, indicating that this sample is closer to optimal doping or contains less nanoscale clustering of $A$-site dopants\cite{Shibata2002} than most samples.

Figure~\ref{Fig2} shows spin wave excitations along the $\Gamma-M$ and $X-M$ directions [$\vec{q}=(\xi,\xi,0$) and $\vec{q}=(0.5,\xi,0$), respectively], measured with $E_i$=120~meV. The higher incident energy yielded a greater dynamic range but coarser energy resolution.  The spin waves at these higher energies are also significantly broadened.  The intrinsic FWHM widths of the spin wave excitations, after deconvolution with the instrumental resolution, are shown in Fig.~\ref{Fig3}(a) as a function of the spin wave excitation energy for reduced wave vectors along three high-symmetry directions: $\vec{q}=(\xi,0,0$)  with $0.1\leq\xi\leq0.5$, $\vec{q}=(\xi,\xi,0$) with $0.08\leq\xi\leq0.29$, and $\vec{q}=(\xi,0.5,0$) with $0\leq\xi\leq0.21$.  (Data points with a negative linewidth represent scans where the measured width was slightly less than the calculated instrumental energy resolution.) Because the ARCS energy resolution is fairly broad, we have also included a few data points with small $q$ using the BT-7 spectrometer in a high-resolution configuration. The dashed line, a guide to the eye, represents zero width (i.e., resolution limited) for excitations at an energy of less than 15~meV and increases linearly with a slope of 1 for excitation energies above 15~meV.  This damping is comparable in scale to that reported by Ye \emph{et al.}\cite{Ye2007} but shows a distinct inflection at $\approx$15~meV; this damping is significantly larger than the $\Gamma=0.24\omega$ reported by Moussa \emph{et al.}\cite{Moussa2007}.  Figure~\ref{Fig3}(b) shows the same data plotted against reduced wave vector magnitude.  In this plot the data along the three directions do not coincide, emphasizing the importance of the excitation energy in determining the linewidth.

Figure~\ref{Fig4} shows example energy scans through the spin waves at $\vec{q}=$($\xi$, 0, 0) with $\xi$=0.14, 0.20, and 0.38. The $\xi$=0.14 and $\xi$=0.20 data were measured on BT-7 while the $\xi$=0.38 data were measured on ARCS with $E_i$=50~meV.  Fits are shown through each data set and horizontal bars represent the full width at half maximum of the instrumental energy resolution, determined by projecting the convolution of the instrumental resolution ellipsoid with the dispersion curve onto the energy axis.  The two spin waves at lower $q$ (with excitation energies of 7.0~meV and 12.5~meV) are resolution limited, consistent with intrinsic linewidths that remain small for excitation energies up to $\approx$15~meV. The raw data for the peak at $\hbar\omega$=7.0~meV fit well to a FWHM of 1.81$\pm$0.16~meV while the instrumental resolution for this peak was calculated to be slightly larger at 2.19$\pm$0.11~meV, consistent only with a small intrinsic linewidth. The raw data for the peak at $\hbar\omega$=12.5~meV fit with a FWHM of 3.88$\pm$0.29~meV while the instrumental resolution was again calculated to be slightly larger at 4.16$\pm$0.20~meV.  An upper bound to intrinsic linewidth of this peak is estimated by shifting the measured linewidth and instrumental resolution of this peak up and down, respectively, by one standard deviation each.  This places an upper bound on any intrinsic linewidth of the spin wave at $\hbar\omega$=12.5~meV as $\Gamma<$1.31~meV. The higher $q$ spin wave, at an excitation energy of 24.2~meV, is significantly broader than resolution with a fit yielding an intrinsic linewidth of $\Gamma$=8.45$\pm$0.47~meV.

\begin{figure}
\centering
\includegraphics[width=8.3cm]{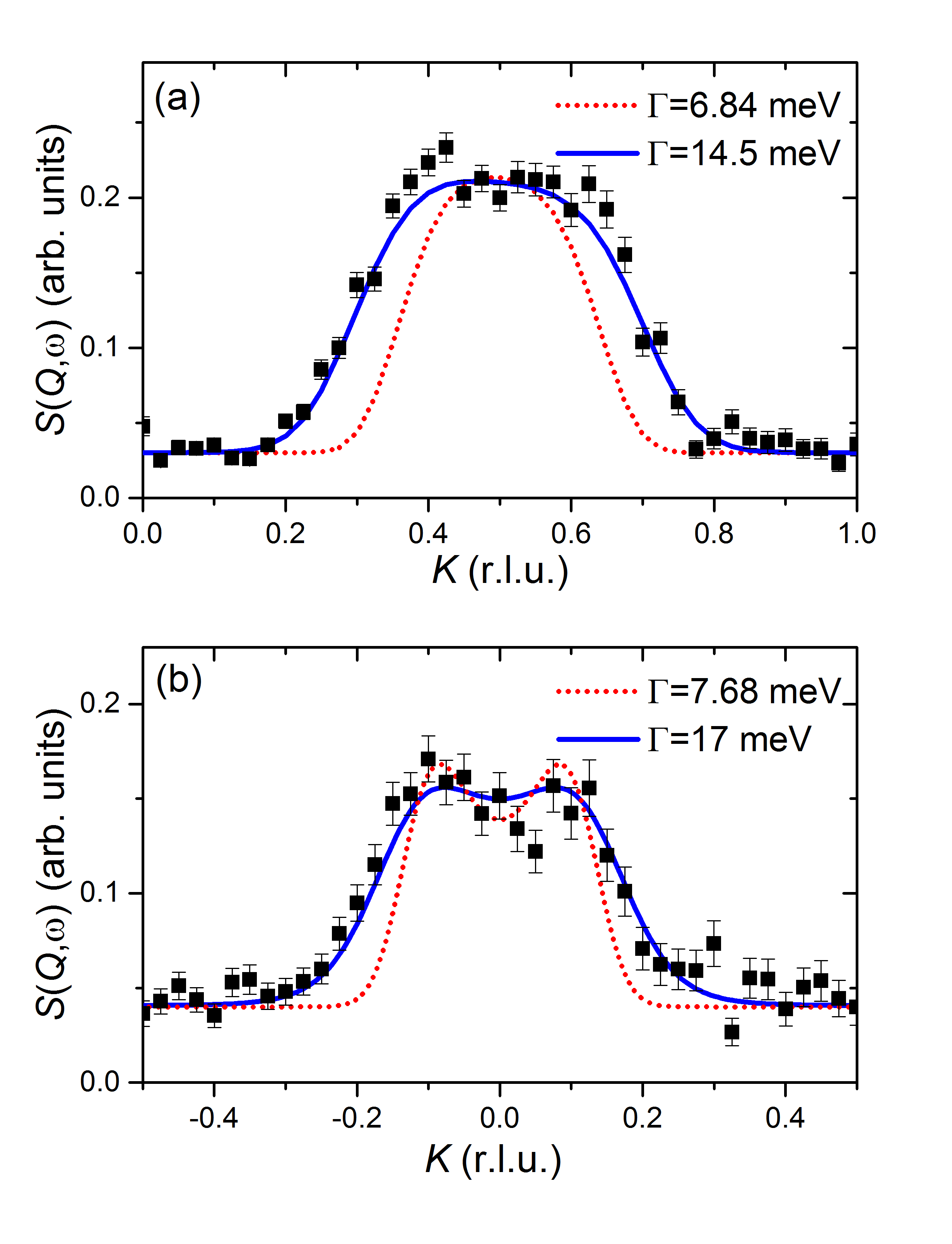} \vspace{-5mm}
\caption{(a) A scan along the (1,$K$,0) direction for 0.95$\leq H \leq$1.05, $-0.1\leq L \leq0.1$, and 27~meV$\leq \hbar\omega \leq$30~meV.  The dotted curve represents a simulation with a spin wave intrinsic width of 6.84~meV,  while the solid curve is for a width of 14.5~meV. (b)  A scan along the (1.5,$K$,0) direction for 1.45$\leq H \leq$1.55, $-0.1\leq L \leq0.1$, and 30~meV$\leq \hbar\omega \leq$34~meV.  The dotted curve represents a simulation with a spin wave intrinsic width of 7.68~meV,  while the solid curve is for a width of 17.0~meV.} \vspace{-2mm}
\label{Fig6}
\end{figure}

Figure~\ref{Fig5} shows the spin wave data within the ($HK0$)~scattering plane alongside a simulation for several energy ranges.  The simulation, utilizing the spin wave dispersion and damping described above, captures the observed data quite well both in terms of the relative intensity and width of the spectrum.  Figure~\ref{Fig6}(a) shows the data for a scan along ($1,K,0$) for energy transfers between 27~meV and 30~meV (the second set of data from the right in Fig.~\ref{Fig5}).  The data are plotted alongside simulations, normalized to match the data at $K$=0.5, of the spin wave scattering.  Simulations are included for intrinsic spin wave widths of both $\Gamma$=6.84~meV (as would be expected for $\Gamma=0.24\omega$) and $\Gamma=14.5$~meV (consistent with our results).  It is clear that a model where the spin wave excitations are not anomalously broadened fails to accurately capture the data in this energy range.  Because the simulations utilize the spin wave dispersion this unambiguously identifies the observed damping with the spin waves themselves and not as an artifact of nearby phonon modes. Figure~\ref{Fig6}(b) shows a scan along ($1.5,K,0$) for energy transfers between 30~meV and 34~meV with similar simulations.  [This scan is symmetry equivalent to a scan through the same reciprocal space point as in Fig.~\ref{Fig6}(a) but in a perpendicular direction.] These data are also consistent with the proposed broadening and inconsistent with a broadening of only $\Gamma=0.24\omega$.

\begin{figure}
\centering
\includegraphics[width=8.3cm]{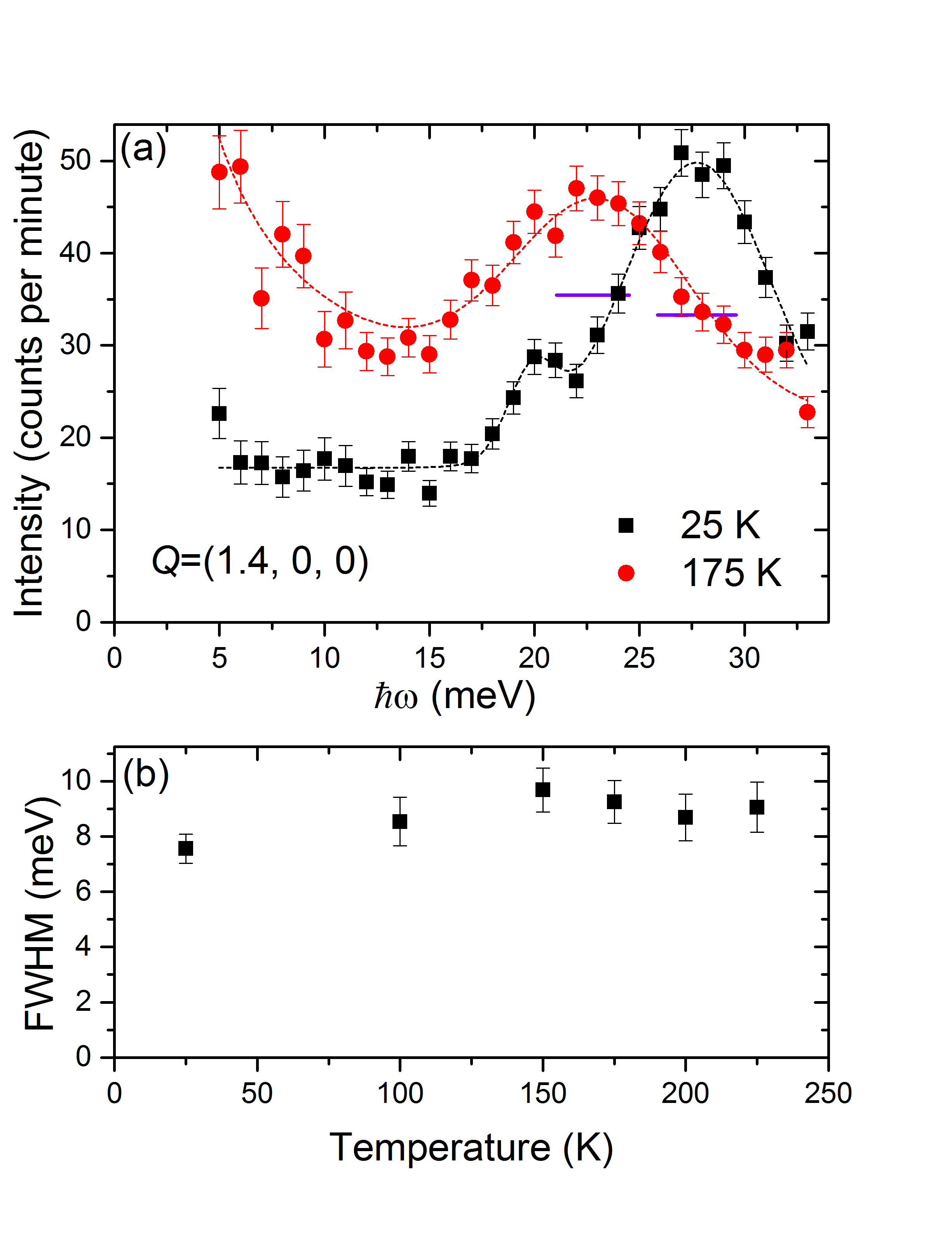} \vspace{-5mm}
\caption{(a) Spin waves measured at $\vec{Q}=(1.4,0,0)$ on the BT-7 spectrometer at $T$=25~K and 175~K. The horizontal lines represent the instrumental resolution(FWHM). (b) The temperature dependence of widths fit from these data.} \vspace{-2mm}
\label{Fig7}
\end{figure}

Figure~\ref{Fig7}(a) displays data from the BT-7 triple-axis spectrometer, showing spin wave excitations at $\vec{Q}$=($1.4,0,0$) at both $T$=25~K and $T$=175~K.  In the low temperature scan the optical phonon branch is visible at around 20~meV; even at this temperature the spin wave is broadened beyond instrumental resolution.  The evolution of this spin wave was tracked with increasing temperature from 25~K to 225~K. For the fits to the higher-temperature data the phonon position and width were fixed to the low temperature values but the intensity was left as a fit parameter. The excitation weakens and softens with increasing temperature; further, a broad quasielastic signal appears at elevated temperatures as spectral weight is shifted into the spin diffusive central component.  The temperature dependence of the width of the spin wave is shown in Fig.~\ref{Fig7}(b).  The width increases slightly while warming to 150~K, but is broad throughout the entire temperature range.

\section{III. Discussion}

From the data presented here, it is clear that this $x$=0.3~LCMO sample displays an anomalous damping for spin wave excitations at energies exceeding $\approx$15~meV.  Earlier reports of a similar broadening were ascribed\cite{Dai2000} to spin wave-phonon scattering, though most experimental evidence points against strong spin wave-phonon coupling.  Scattering of spin waves by electrons\cite{Golosov2000,Golosov2010,Solontsov2013} and orbital excitations\cite{Krivenko2004} have also been theorized to lead to spin wave damping.  Direct comparison to the theory\cite{Harris} of magnon-magnon interaction is difficult in the regime where $\hbar\omega\approx k_BT$, but at higher energies this material clearly does not display the $\Gamma\propto q^3$ behavior found at lower temperatures in materials where magnon-magnon scattering is the primary cause of broadening\cite{Bohn}. Additionally, the role that randomness or disorder can play in the magnetism and charge transport of manganites has been studied for decades\cite{Sheng1997,Li1997} and might play a role in damping.  Structural disorder, arising from the differing ionic radii of the $A$-site cations, could lead to a variation of Mn-O-Mn bond angles and therefore increase the bandwidth of the ferromagnetic exchange interaction and lead to spin wave damping\cite{Motome2005}.  However this structural disorder, characterized by the variance in $A$-site ionic radius ($\sigma_A^2$)\cite{Rodriguez1996}, should be much smaller in LCMO compared to other manganites given that the La$^{3+}$  and Ca$^{2+}$  ionic radii are quite similar\cite{Shannon1976}.  None of these scenarios are sufficient to describe the observed damping. 

The role of phase separation\cite{Dagotto2001} in CMR manganites has also attracted a great deal of attention.  Results from scanning tunneling microscopy\cite{Fath1999} and electron microscopy\cite{Uehara1999} have reported CMR as a percolation effect of metallic regions, with insulating regions persisting well below $T_c$.  Similarly, even lightly doped samples that do not display a metallic ground state do show evidence of nanoscale ferromagnetic or orbitally ordered domains\cite{Hennion1998,Hennion2005}.  While structural disorder due to differing $A$-site ionic radii should be quite small, this material displays local Jahn-Teller structural distortions surrounding polarons that become self-trapped above $T_c$ when a conduction electron localizes around a Mn$^{3+}$ ion\cite{Lynn2007}.  If phase separation in $x$=0.3~LCMO led to nanoscale regions with dynamic Jahn-Teller distortions even in the ferromagnetic state, the increased bandwidth in the ferromagnetic exchange interaction due to bond length variation surrounding the distortions could explain the observed spin wave broadening.  A similar dynamic phase separation has been invoked to explain the damping of an optical Jahn-Teller phonon mode (near 70~meV) at temperatures approaching $T_c$\cite{Zhang2001}.  Long-wavelength excitations would average over several domains and not broaden as significantly.  The broadening along both the [100] and [110] directions begins near $q=0.35$~{\AA}$^{-1}$, suggesting a typical domain size at $T$=100~K of $\approx$18~{\AA}.  This length scale is comparable to that of both polaron correlations\cite{Adams2000,Dai2000prl} and magnetic clusters\cite{deTeresa} above $T_c$.  The broadening along various high-symmetry directions is dependent on the excitation energy as opposed to the magnitude of the reduced wave vector (Fig.~\ref{Fig3}), which could be related to the hopping nature of the polarons\cite{Helton2014}.  The relatively slight temperature dependence of the spin wave damping near the Brillouin zone boundary (Fig.~6) suggests that the spatial extent of the phase separated regions does not change strongly with temperature. This is notably comparable to the temperature dependence of the widths of polaron scattering measured in elastic scans in LCMO\cite{Lynn2007} and inelastic scans in La$_{0.7}$Sr$_{0.3}$MnO$_3$ and La$_{0.7}$Ba$_{0.3}$MnO$_3$\cite{Chen} (which in LSMO have also been attributed to the damping of acoustic phonons\cite{Maschek}), though distinct from the temperature dependence of the size of dynamically fluctuating phase separated regions previously posited in LCMO\cite{Zhang2001}.

\section{IV. Summary}

In summary, neutron spectroscopy measurements on a sample of colossal magnetoresistive La$_{0.7}$Ca$_{0.3}$MnO$_3$ close to optimal doping ($T_c=$257~K) reveal spin wave excitations that are anomalously damped, with intrinsic widths that rise sharply for excitation energies exceeding $\approx$15~meV and reach an intrinsic FWHM of at least 20~meV.  This width is inherent to the spin waves and cannot be explained as an artifact of a nearby phonon mode. The onset of broadening corresponds to a length scale of about 18~{\AA}.  This is comparable to size estimates for the nanoscale phase inhomogeneities of insulating or metallic domains that yield percolative colossal magnetoresistance.  We suggest that insulating nanoscale regions featuring dynamic Jahn-Teller distortions persisting at temperatures well below $T_c$ increase the bandwidth of ferromagnetic exchange interactions due to bond angle distortions, damping short-wavelength spin waves.  Spin wave damping has been observed in a variety of manganite materials, with varying details that may defy a single explanation.  However, this work clearly shows anomalously damped spin wave excitations in $x$=0.3~LCMO, where chemical pressure is small and unlikely to have a significant effect.

\section*{ACKNOWLEDGEMENTS}
We thank J. Fernandez-Baca for helpful discussions.  J.S.H acknowledges partial support from the Office of Naval Research Mathematics, Computer and Information Sciences Division through the Naval Academy Research Council. D.A.S acknowledges the Ministry of Education and Science of the Russian Federation (Grant No. 14.Y26.31.0005). A portion of this research used resources at the Spallation Neutron Source, a DOE Office of Science User Facility operated by the Oak Ridge National Laboratory.

emails: $\ast$ helton@usna.edu, $\dag$ jeffrey.lynn@nist.gov\\

\bibliography{LCMOWidths}

\end{document}